\documentclass[prb, showpacs,showkeys,preprint]{revtex4}

\usepackage{graphicx}

\begin{document}

\title{Geometrical derivation of the Boltzmann factor}

\author{Ricardo L\'opez-Ruiz}
\email{rilopez@unizar.es}
\affiliation{
DIIS and BIFI, Facultad de Ciencias, \\
Universidad de Zaragoza, E-50009 Zaragoza, Spain}

\author{Jaime Sa\~nudo }
\email{jsr@unex.es}
\affiliation{
Departamento de F\'isica, Facultad de Ciencias, \\
Universidad de Extremadura, E-06071 Badajoz, Spain}

\author{Xavier Calbet}
\email{xcalbet@googlemail.es}
\affiliation{
Instituto de Astrof{\'\i}sica de Canarias, \\
V{\'\i}a L\'actea, s/n, 
E-38200 La Laguna, Tenerife, Spain}

\date{\today}

\begin{abstract}
We show that the Boltzmann factor has a geometrical origin.
Its derivation follows from the microcanonical picture. 
The Maxwell-Boltzmann distribution or the wealth distribution
in human society are some direct applications of this new interpretation.
\end{abstract}


\maketitle

The Maxwellian distribution was derived in Ref. \onlinecite{lopez2007}
by geometrical arguments over an $N$-sphere.
Following this insight, here we explain the origin of the Boltzmann 
factor by means of the geometrical properties of an $N$-hyperplane.
This alternative view can be thought as an a priori theoretical argument
that explains why the probability of a magnitude, constrained by a conservation law, 
is given by the exponential distribution when the ensemble or system over which that magnitude 
is measured is in {\it thermodinamical equilibrium}.

By thermodinamical equilibrium we mean the macroscopic stationary regime
where all the microscopic accessible states of the system are equiprobable.
This implies that the deterministic or random time evolution of the system
verifies the ergodic hypothesis. This has been proved in very few systems \cite{sinai}, and hence, 
in general, it is supposed as an hypothesis in classical statistical mechanics \cite{huang}.
If the system do not explore its own phase space with equiprobable results for all 
the accessible states then the system exhibits some kind of ergodicity breaking
and the final distribution will not be the exponential distribution. We can call it
a {\it non-Boltzmann equilibrium}.   

Without loss of generality, here we obtain analytically the Boltzmann factor
(or Maxwell-Boltzmann distribution) in a model where the ergodic hypothesis 
has been computationally proved. The method was developed in 
Ref. \onlinecite{lopez2007} and it is grounded on geometrical arguments.  

The model was proposed by Dragulescu and Yakovenko \cite{yakovenko1} 
in order to explain the distribution of wealth in human society.
Thus it is known that the incomes of 90\% of the population in western societies
can be fitted by a exponential distribution.\cite{yakovenko2} 
Supposing initially equity in a set of many agents, that is, all of them have
the same initial quantity of money, the model establishes random binary interactions 
in which agents exchange money but conserve its total amount. 
It is computationally found that the system asymptotically tends toward 
the exponential distribution. This means that, in this case,
the evolution mechanisms proposed by Dragulescu and Yakovenko for 
the money exchange among agents verify the ergodic hypothesis 
and, then, its final distribution is the exponential distribution.
(Let us say that other non-Boltzmann random mechanisms for the money exchange
are also proposed in Ref. \onlinecite{yakovenko1}. These mechanisms imply the
breaking of the ergodic hypothesis and therefore the asymptotic evolution of the system
toward non-Boltzmann equilibria).

We start by assuming $N$ agents, each one with coordinate (money) $x_i$, $i=1,\ldots,N$, 
with $x_i\geq 0$, and a total amount of money $E$ that is conserved:
\begin{equation}
x_1+x_2+\cdots +x_{N-1}+x_N = E.
\label{eq-E}
\end{equation} 
Under the evolution rules proposed in Ref. \onlinecite{yakovenko1},
this isolated system evolves on the positive part of an equilateral 
$N$-hyperplane (i.e., the surface formed for all those points with $x_i\geq 0$, for all $i$). 
The formula for the surface area $S_N(E)$ of an equilateral $N$-hyperplane of side $E$ is
\begin{equation}
S_N(E) = {\sqrt{N}\over (N-1)!}\;E^{N-1}.
\label{eq-S_n}
\end{equation}
(See appendix A for the derivation of this formula). If the ergodic hypothesis is 
assumed, each state of the microcanonical ensemble (that is, each point
on the $N$-hyperplane) is equiprobable, then the probability $f(x_i)dx_i$ of finding 
the agent $i$ with money $x_i$ is proportional to the 
surface area formed by all the points on the $N$-hyperplane having the $i$th-coordinate 
equal to $x_i$. Our objective is to show that $f(x_i)$ is the Boltzmann factor
(or the Maxwell-Bolztamnn distribution), with the normalization condition
\begin{equation}
\int_{0}^Ef(x_i)dx_i = 1.
\label{eq-p_n}
\end{equation}

If the $i$th agent has coordinate $x_i$, the $N-1$ remaining agents 
share the money $E-x_i$ on the $(N-1)$-hyperplane
\begin{equation}
x_1+x_2 \cdots +x_{i-1} + x_{i+1} \cdots +x_N= E-x_i,
\label{eq-E1}
\end{equation} 
whose surface area is $S_{N-1}(E-x_i)$. 
If we define the coordinate $\theta_N$ (see appendix A) as satisfying
\begin{equation}
\sin\theta_N = \sqrt{N-1 \over N},
\label{eq-theta}
\end{equation}
it can be easily proved that
\begin{equation}
S_N(E) = \!\int_{0}^{E}\!S_{N-1}(E-x_i) {dx_i \over \sin\theta_N}.
\label{eq-theta1}
\end{equation}

Hence, the surface area of the $N$-hyperplane for which the $i$th coordinate is
between $x_i$ and $x_i+dx_i$ is $S_{N-1}(E-x_i)dx_i/\sin\theta_N$.
We rewrite the surface area as a function of $x_i$, 
normalize it to satisfy Eq.~(\ref{eq-p_n}), and obtain
\begin{equation}
f(x_i) = {1\over S_N(E)}
{S_{N-1}(E-x_i)\over \sin\theta_N},
\label{eq-f_n}
\end{equation}
whose final form, after some calculation is
\begin{equation}
f(x_i) = (N-1)E^{-1}\Big(1-{x_i\over E} \Big)^{N-2},
\label{eq-mm}
\end{equation}
If we call $\epsilon$ the mean wealth per agent, 
$E=N\epsilon$, then in the limit of large $N$ 
we have
\begin{equation}
\lim_{N\gg 1}\left(1-{x_i\over E}\right)^{N-2}
\simeq e^{-{x_i/\epsilon}}.
\label{eq-ee}
\end{equation}
The Boltzmann factor $e^{-{x_i/\epsilon}}$ is found 
when $N\gg 1$ but, even for small $N$, it can be a good approximation 
for agents with low wealth. After substituting Eq.~(\ref{eq-ee})
into Eq.~(\ref{eq-mm}), we obtain the Maxwell-Boltzmann distribution 
in the asymptotic regime $N\rightarrow\infty$ (which also implies $E\rightarrow\infty$):
\begin{equation}
f(x)dx = {1\over \epsilon}\,e^{-{x/\epsilon}}dx,
\label{eq-gauss}
\end{equation}
where the index $i$ has been removed because the distribution is the same for each agent, 
and thus the wealth distribution can be obtained by averaging over all the agents. 

Depending on the physical situation the mean wealth per agent $\epsilon$
takes different expressions and interpretations. 
For instance, in the case of an isolated one-dimensional gas,
the $x_i$ variable represents the energy of the particle $i$,
thus we can calculate the dependence of $\epsilon$ on the temperature, which
in the microcanonical ensemble is defined by differentiating the entropy
with respect to the energy. The entropy can be written as $S=-kN\!\int_{-\infty} 
^{\infty} f(x)\ln f(x)\,dx$, where $f(x)$ is given by Eq.~(\ref{eq-gauss})
and $k$ is the Boltzmann constant. 
If we recall that $\epsilon=E/N$, we obtain
\begin{equation}
S(E)= kN\ln\left({E\over N} \right) + kN.
\end{equation}
The calculation of the temperature $T$ gives
\begin{equation}
T^{-1}= \left({\partial S\over \partial E} \right)_N = {kN\over E} = {k\over \epsilon}.
\end{equation}
Thus $\epsilon=kT$, and the Maxwell-Boltzmann distribution 
is obtained as it is usually given in the literature:
\begin{equation}
f(x)dx = {1\over kT}\,e^{-x/kT}dx.
\end{equation}

In conclusion, we have shown that the Boltzmann factor describes 
the general statistical behavior of each small part 
of a multi-component system whose components or parts are given
by a set of random variables that satisfy a conservation law (condition (\ref{eq-E})),
and that run in an equiprobable manner over all the states of its own phase space.


\setcounter{equation}{0}
\section*{APPENDIX A: \underline{Formula for the surface of an equilateral $N$-hyperplane}}

Here we derive the formula for the calculation of the surface $S_N(E)$ of an equilateral 
$N$-hyperplane, $\Pi_N$, embedded in $\Re^N$ and formed by all the points $(x_1,x_2,\ldots,x_N)$,
with $x_i\geq 0$ for all $i=1,2,\ldots,N$, verifying the equation 
\begin{equation}
x_1+x_2+\cdots +x_{N-1}+x_N = E,
\label{eq-E2}
\end{equation}
with $E$ a real constant.

First we define the angle $\theta_N$ as follows.
The unitary perpendicular vector to the hyperplane $\Pi_N$ in $\Re^N$, $\omega_\perp$, 
has coordinates ${1\over \sqrt{N}}(1,1,\ldots,1,1)$. The unitary vector in the $x_N$-direction, 
$\upsilon_{x_N}$, has coordinates $(0,0,\ldots,0,1)$. The projection of 
$\omega_\perp$ over $\upsilon_{x_N}$ is given by the scalar product of both vectors:
\begin{equation}
\cos\theta_N = \omega_\perp \cdot \upsilon_{x_N} = {1 \over \sqrt{N}}\;,
\end{equation}
that defines the angle $\theta_N$. Let us observe the curious property of this angle:
\begin{equation}
{\cos\theta_N \over \cos\theta_{N-1}} = \sin\theta_N.
\end{equation}

The surface $S_N(E)$ in $\Re^N$ is the measure of a set of points with 
$N-1$ dimensions, just the set of all those points $(x_1,x_2,\ldots,x_N)$, with $x_i\geq 0$, 
verifying Eq. (\ref{eq-E2}). The one-to-one projection of this set of points on the $x_N$-direction
generates a set of points of $N-1$ dimensions in $\Re^{N-1}$, just the set of all those points
in $\Re^{N-1}$ satisfying the inequality:
\begin{equation}
x_1+x_2+\cdots +x_{N-1} \leq E.
\label{eq-E3}
\end{equation}
The volume of this set of points $V_{N-1}(E)$ is the volume  
of an $(N-1)$-dimensional pyramid formed by $N$ vertices linked by
$N-1$ perpendicular sides of length $E$. 
It is straightforward to see that the formula for this volume is:
\begin{equation}
V_{N-1}(E) = {E^{N-1} \over (N-1)!}.
\end{equation}
Moreover, $V_{N-1}(E)$ and $S_N(E)$ can be calculated explicitly
when $N=1,2,3,4$. They verify the relationship:
\begin{equation}
V_{N-1}(E) = S_N(E)\cdot \cos\theta_N.
\end{equation}
This expresses an exact connection between the surface $S_N(E)$ 
of the $N$-hyperplane in $\Re^{N}$ and its projection
in the $x_N$-direction over the $(N-1)$-dimensional volume $V_{N-1}(E)$ in $\Re^{N-1}$
for those low dimensional cases. Extrapolating this behavior for all $N$, 
we conclude that 
\begin{equation}
S_N(E) = {V_{N-1}(E) \over \cos\theta_N}={\sqrt{N}\over (N-1)!}\;E^{N-1}.
\label{ver}
\end{equation}
as it has been used in Eq. (\ref{eq-S_n}) of the main text.

Although we can not claim the mathematical proof of formula (\ref{ver}),
the Boltzmann factor is exactly obtained in the limit $N\gg1$
after taking this formula as starting point of all the line of reasoning 
developed in the main text. This result can be used like {\it a posteriori} argument 
that proves the correctness of expression (\ref{ver}).

\setcounter{equation}{0}
\section*{APPENDIX B: \underline{A possible generalization: an open problem}}
In this work, we have shown that an ensemble of positive variables $(x_1,x_2,\ldots,x_N)$ verifying 
\begin{equation}
x_1+x_2+\cdots +x_{N-1}+x_N = E,
\label{eq-En}
\end{equation}
with an adequate mechanism assuring the ergodic hypothesis, i.e.,
the equiprobability of all the possible states $(x_1,x_2,\ldots,x_N)$
on the hypersurface of the phase space,
presents the exponential distribution of the generic variable $x$ over the 
ensemble when $N\rightarrow\infty$,
\begin{equation}
f(x)dx \sim \epsilon^{-1}\,e^{-{x/\epsilon}}dx,
\label{eq-gauss1}
\end{equation}
with $E=N\epsilon$ and $\epsilon$ the mean value of the real constant $E$
over the collectivity.  

In the work with Ref. \onlinecite{lopez2007}, 
it was shown that an ensemble of positive variables $(x_1,x_2,\ldots,x_N)$ verifying 
\begin{equation}
x_1^2+x_2^2+\cdots +x_{N-1}^2+x_N^2 = E,
\label{eq-Em}
\end{equation}
with an adequate mechanism assuring the ergodic hypothesis, i.e.,
the equiprobability of all the possible states $(x_1,x_2,\ldots,x_N)$
on the hypersurface of the phase space,
presents the Gaussian distribution of the generic variable $x$ over the 
ensemble when $N\rightarrow\infty$,
\begin{equation}
f(x)dx \sim \epsilon^{-1/2}\,e^{-{x^2/2\epsilon}}dx,
\label{eq-gauss2}
\end{equation}
with $E=N\epsilon$ and $\epsilon$ the mean value of the real constant $E$
over the collectivity.  

The general question that we want to bring to the reader is the following.
Let $b$ be a real constant. 
If we have a set of positive variables $(x_1,x_2,\ldots,x_N)$ verifying 
\begin{equation}
x_1^b+x_2^b+\cdots +x_{N-1}^b+x_N^b = E
\label{eq-Ek}
\end{equation}
with an adequate mechanism assuring the ergodic hypothesis, i.e.,
the equiprobability of all the possible states $(x_1,x_2,\ldots,x_N)$
on the hypersurface of the phase space,
will we have for the generic variable $x$ the distribution
\begin{equation}
f(x)dx \sim \epsilon^{-1/b}\,e^{-{x^b/b\epsilon}}dx,
\label{eq-gaussn}
\end{equation}
when we average over the ensemble in the limit $N\rightarrow\infty$?. 
The answer to this question, as far as we know, is an open problem.

\end{document}